\documentclass[aip,jcp,twocolumn,superscriptaddress,10pt]{revtex4-1}
\usepackage{color}
\usepackage{fancyhdr} 
\usepackage{varwidth} 
\usepackage{graphicx,subfigure}  
\usepackage{bm} 
\usepackage{listings}
\usepackage{enumerate}%
\usepackage[colorlinks,citecolor=blue,linkcolor=blue]{hyperref}
\usepackage{amsmath}
\usepackage{threeparttable}
\usepackage{graphicx}
\usepackage{amssymb}
\usepackage{amsthm}
\usepackage{rotating}
\usepackage{multirow}

\makeatletter
\def\@email#1#2{%
 \endgroup
 \patchcmd{\titleblock@produce}
  {\frontmatter@RRAPformat}
  {\frontmatter@RRAPformat{\produce@RRAP{*#1\href{mailto:#2}{#2}}}\frontmatter@RRAPformat}
  {}{}
}%
\makeatother

\begin{document}
\title{Local collective dynamics at equilibrium BCC crystal-melt interfaces}

\author{Xin Zhang}
\affiliation{State Key Laboratory of Precision Spectroscopy, School of Physics and Electronic Science, East China Normal University, Shanghai 200241, China}
\author{Wenliang Lu}
\affiliation{School of Science, Changzhou Institute of Technology, Changzhou, Jiangsu 213032, China}
\author{Zun Liang}
\affiliation{State Key Laboratory of Precision Spectroscopy, School of Physics and Electronic Science, East China Normal University, Shanghai 200241, China}
\author{Yashen Wang}
\affiliation{State Key Laboratory of Precision Spectroscopy, School of Physics and Electronic Science, East China Normal University, Shanghai 200241, China}
\author{Songtai Lv}
\affiliation{State Key Laboratory of Precision Spectroscopy, School of Physics and Electronic Science, East China Normal University, Shanghai 200241, China}
\author{Hongtao Liang}
\affiliation{State Key Laboratory of Precision Spectroscopy, School of Physics and Electronic Science, East China Normal University, Shanghai 200241, China}
\author{Brian B. Laird}
\affiliation{Department of Chemistry, University of Kansas, Lawrence, KS 66045, USA}
\author{Yang Yang}
\thanks{yyang@phy.ecnu.edu.cn}
\affiliation{State Key Laboratory of Precision Spectroscopy, School of Physics and Electronic Science, East China Normal University, Shanghai 200241, China}
\affiliation{Chongqing Institute of East China Normal University, Chongqing 401120, China}
\date{\today}

\begin{abstract}

We present a classical molecular-dynamics study of the collective dynamical properties of the coexisting liquid phase at equilibrium body-centered cubic (BCC) Fe crystal-melt interfaces. For the three interfacial orientations (100), (110), and (111), the collective dynamics are characterized through the calculation of the intermediate scattering functions,  dynamical structure factors and density relaxation times in a sequential local region of interest. An anisotropic speed up of the collective dynamics in all three BCC crystal-melt interfacial orientations is observed. This trend differs significantly different from the previously observed slowing down of the local collective dynamics at the liquid-vapor interface [Acta Mater 2020;198:281]. Examining the interfacial density relaxation times, we revisit the validity of the recently developed time-dependent Ginzburg-Landau (TDGL) theory for the solidification crystal-melt interface kinetic coefficients, resulting in excellent agreement with  both the magnitude and the kinetic anisotropy of the CMI kinetic coefficients measured from the non-equilibrium MD simulations.

\end{abstract}

\maketitle

\section{introduction}

Quantitatively predicting the anisotropic crystal-melt interface (CMI) kinetics with given interface orientation $\hat{n}$ has been a long term goal in the field of controlling solidification or crystal growth, for example through the kinetic coefficient $\mu_{\hat{n}}$, defined as the constant of proportionality between the steady-state solidification interface velocity $V_{\hat{n}}$ and the degree of the interface undercooling $\Delta T$. The most recent advancement in  quantitative prediction of near-equilibrium CMI kinetics is the so-called time-dependent Ginzburg-Landau (TDGL) theory\cite{Wu15,Xu20} of solidification kinetics. Compared with previous theories or models, the TDGL theory significantly improves the prediction of the kinetic coefficient, predicting a correct anisotropy sequence, while remaining free of undetermined fitting parameters. This theory has also demonstrated unique potential for revealing the nature of the kinetic anisotropy\cite{Wang22} (see in Appendix \ \ref{sec:ap1}). However, recent validation studies of the derived TDGL theory suggest that a 15\% to 20\% discrepancy between the molecular-dynamics (MD) simulation measurements\cite{Wu15,Xu20,Wang22} $\mu^{\mathrm{MD}}_{\hat{n}}$ and the TDGL predictions $\mu^{\mathrm{GL}}_{\hat{n}}$ exists in different CMIs, and not in a fixed fashion as material varies.

The clarification of the above mentioned discrepancies between $\mu^{\mathrm{MD}}_{\hat{n}}$ and $\mu^{\mathrm{GL}}_{\hat{n}}$ is needed to understand  two fundamental issues. First, one needs a quantitative predictive theory that is sufficiently accurate in order to explain why universal kinetic anisotropy trend that holds in the FCC CMIs over a range of different materials\cite{Asta09}, does not extend to BCC CMI systems\cite{Hoyt06}. Second, an accurate kinetic theory is critical to unifying  kinetics models based on transition state theory (such as the Wilson-Frenkel model\cite{Wilson1900,Frenkel1932,Mendelev10}, BGJ model\cite{Broughton82}, or the local structure-dependent model proposed by Freitas et al.\cite{Rodrigo20}) with the quantitative TDGL theory, in order to terminate the debated recognitions of the limiting mechanism of the solidification.

The authors have pointed out clues that may shed light on the origin of the discrepancies between $\mu^{\mathrm{MD}}_{\hat{n}}$ and $\mu^{\mathrm{GL}}_{\hat{n}}$\cite{Xu20,Wang22}. Specifically, the appropriateness of using the bulk value of a time-scale parameter (i.e., the density relaxation time associated with the collective dynamics of the liquid atoms), in the TDGL theory\cite{Wu15} prediction of the anisotropic $\mu^{\mathrm{GL}}_{\hat{n}}$ (i.e., see Eq.\ref{MunM} and \ref{AnM} in Appendix \ref{sec:ap1}), has been questioned. The authors raised a hypothesis that the dynamic time scale due to density fluctuation at the CMI holds the same value in the bulk melt phase is inappropriate, instead, such dynamic time scale should have anisotropic and material-dependent values at different oriented CMIs. Nevertheless, until the current study, the validity of this hypothesis has not been verified.

Studies of interfacial local collective dynamics have been carried out in the liquid-vapor surface systems through experiments\cite{Reichert07,Wehinger11} and atomistic simulations\cite{Iarlori89,Rio20}. Reichert et al. employed  low angle beam X-ray scattering\cite{Reichert07} 
 to explore the collective dynamics of the subsurface region within a few nanometers of the surface of of molten metals They observed a slowing down in the collective dynamics and a larger longitudinal viscosity within this subsurface region compared to the bulk melt phase. Following this experimental study, del Rio et al.\cite{Rio20} calculated the collective dynamical properties of this same system at varying depths, obtaining excellent agreement with the experimental data. In addition, they were able to compute the properties much closer to the surface (around one to two atomic layers) and have discovered that, at these shallower depths, the properties drastically differ from those deeper in the slab.\par
To the best of our knowledge, there is no inelastic neutron or X-ray scattering
experimental study on the collective dynamics of the local liquids at CMIs. In contrast to the liquid-vapor system, the CMI is sandwiched between two condensed phases making experimental study difficult. In addition, the impact of the crystal structure on the adjacent liquid layers may affect both the layering order and the lateral (in-plane) ordering\cite{Kaplan06}. It is unknown whether the algorithm for the calculation of the local collective dynamics proposed by del Rio et al.,\cite{Rio20} remains valid for CMI systems.

This study proposes an improved method for calculating the collective dynamics of the interfacial liquid  through an analysis of the trajectories of equilibrium molecular-dynamics simulations. The improved calculation method as well as the method of del Rio and Gonz\'{a}lez are applied to equilibrium Fe BCC(100), (110), and (111) CMIs, yielding the spatial variations of the intermediate scattering functions (ISFs), the dynamical structure factors (DSFs) and the density relaxation times for a wavenumber corresponding to the BCC principal reciprocal lattice vector (RLV). Instead of the bulk liquid value, the density relaxation times for the interfacial liquids are used for the subsequent applications of the TDGL theory. We demonstrate that the current improved method provides more precise data than the previous method and eliminates the discrepancy between the $\mu^{\mathrm{MD}}_{\hat{n}}$ and $\mu^{\mathrm{GL}}_{\hat{n}}$. This results in in reasonable predictions of both the magnitude and the anisotropy in $\mu$ for the three BCC Fe CMIs. The current computational effort on the CMIs enriches the well-known atomistic simulation-based characterization methodology (e.g., structure, thermodynamics, and dynamical properties\cite{Davidchack98}) for the solid-liquid interface systems. Our study could benefit the understanding of the solidification kinetic anisotropy near or far-from equilibrium\cite{Asta09,Ashkenazy10}, as well as the solidification-melting asymmetry in a certain amount of materials\cite{Tsao86,Celestini02,Hwang19,Wu21}, facilitate the multi-scale modeling of the solidification microstructure evolution (control), as well as the development of the statistical mechanical theories for the prediction of the local collective dynamics of interfacial liquids\cite{Bafile06}.

\section{Computational Methods}

\subsection{MD Simulations}
  
The embedded-atom method potentials for BCC Fe\cite{Mendelev03} employed in the current MD simulations have been used in several previous studies of BCC Fe CMIs. For example, Sun et al.\cite{Sun04} and Gao et al. \cite{Gao10} studied the interfacial free energies and the kinetic coefficients of the Fe CMIs using this EAM potential. In addition, Wu et al. determined the interfacial profiles\cite{Wu06} of the density wave amplitude order parameter in the framework of the Ginzburg-Landau theory of the bcc-melt interfaces and, in a recent work, Lu et al.  calculated the excess interfacial stresses of the BCC Fe CMIs described by this EAM potential\cite{Lu22}. Here we choose this particular potential to revisit the validity of the TDGL theory for the $\mu^{\mathrm{GL}}_{\hat{n}}$, the current results are compared to the $\mu^{\mathrm{MD}}_{\hat{n}}$\cite{Gao10} and the previous TDGL predictions\cite{Wu15} using this potential.

\begin{figure*}[!htb]
\centering
\includegraphics[width=0.9\textwidth]{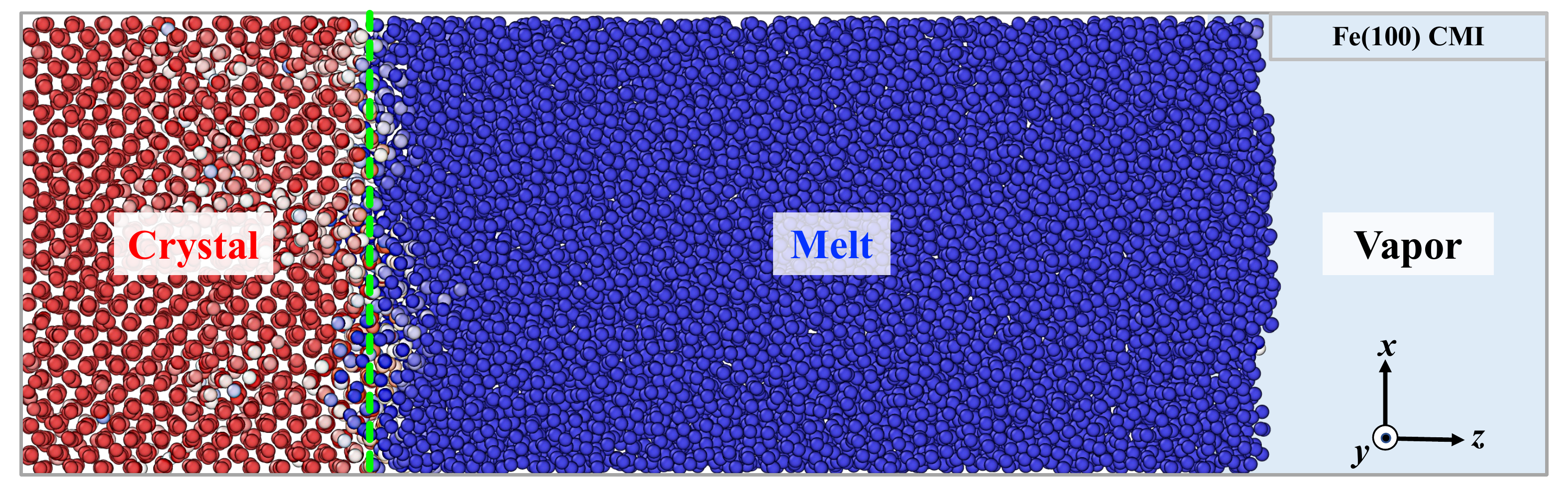}
\caption{Snapshot from an MD simulation of the equilibrium Fe(100) CMI at 1772K ($T_\mathrm{m}$). The crystal phase coexists with the liquid phase, which is exposed to the vapor phase to ensure zero pressure hydrostatic condition along the $z$ axis direction or the CMI normal direction. The Fe atoms are color-coded as crystalline (red) or melt (blue) according to the value of a local structural order parameter $\psi$ \cite{Morris02}.  The values ${\psi} = 1$ and 0 correspond to the perfect crystalline phase and the melt phase, respectively. The dashed line denotes the instantaneous CMI position defined from the atomic order parameter distribution along $z$\cite{Yang11}}.
\label{fig1}
\end{figure*}

We perform the calculations of the local liquid dynamical properties based on the equilibrium MD simulations of three  BCC Fe CMIs, namely, (100), (110), and (111). All the details, procedures, and boundary conditions for constructing the equilibrium CMIs, based on the platform of LAMMPS software\cite{Steve95}, are precisely the same as those reported in the Ref.\onlinecite{Lu22}, including the particle number size (35100, 32340, 34056 atoms for the three CMI orientations) and dimensions of the simulation cell, see in Fig.\ref{fig1}. In comparing with the dynamical properties measured from the interfacial liquids, we carry out an independent simulation of a bulk melt sample containing 2000 Fe atoms, at the melting temperature for this model, $T_\mathrm{m}=1772$K, at pressure $p=1.0$ bar.
Both isothermal-isobaric (constant $NpT$) and canonical (constant $NVT$) MD simulations using the Nos\'e-Hoover thermostat (relaxation time 100 fs) and Anderson barostat (relaxation time  500 fs) are employed in preparing the bulk melt sample using a time-step of 1fs. Note that, analysis with tests of the dependence of collective dynamical properties on thermostat relaxation times are carried out to confirm that the density relaxation times are independent of the value of the adjustable thermostat parameters over an interval spanning three orders of magnitude (10fs, 100fs, 1000fs), the the density relaxation times obtained with different thermostat parameters are scattered by no more than 6\% around an average value, consistent with the statistical uncertainties.

In order to collect samples for the calculation of the time-spatial correlation functions and the equilibrium averages in the equilibrated CMIs and bulk melt systems, we subdivide the simulation into 10 ps blocks. Within each block, 5000 successive MD trajectories (each separated by 2 fs) are recorded for the calculation of the ISFs and DSFs. We use 500 blocks (across 50 ns equilibrium CMIs simulations). This procedure provides 1000 samples (as there are two independent interfaces in each simulation cell) for statistical averaging. In order to ensure independence, each block used  in the statistical averaging is separated in time by at least 50 ps.

\subsection{Calculation of interfacial profiles}

In this study, we examine two types of interfacial profiles. The first type is the mean fine-grained particle number density profile $\rho(z)$, measured for each of the three interfacial orientations, and calculated using the protocol outlined in Ref.\onlinecite{Lu22}. The instantaneous fine-grained density profiles determined from the instantaneous MD trajectory (1ps each), are aligned in $z$ axis to the time-averaged CMI position, defined as $z=0$, for calculating the time-averaged density profile. In which, the instantaneous CMI position is determined from the local structural order parameter $\psi$\cite{Morris02} distribution along $z$\cite{Yang11}, the values ${\psi} = 1$ and 0 correspond to the perfect crystalline phase and the melt phase, respectively. In this way, the calculated density profile is unbroadened by the capillary fluctuations.
Following Rull and Toxvaerd\cite{Rull83}, the maximum of each of the peaks in the density profiles for $z > 0$ [$\rho_\mathrm{max}(z)$] are labeled in order to quantify the asymptotic decay length of stratification for each CMIs, through fitting the $[\rho_\mathrm{max}(z)-\rho_\mathrm{M}]$ via exponential function $\propto \exp[-z/\zeta_{\hat{n}}]$. Here the subscript ``M'' denotes the bulk melt phase, $\zeta_{\hat{n}}$ is the decay distance from the CMI position with given interface orientation $\hat{n}$.

The second type is the coarse-scale profile of the characteristic time-scales (i.e., density wave relaxation times), $\tau_{\hat{n}}(z_m)$, depicting the CMI local collective dynamics. On the liquid side of each CMI orientation ($z>0$), from $z=3$\ \AA \ to $z=18$\ \AA, we report $\tau_{\hat{n}}(z_m)$ results at 8 different positions along $z$ with an interval of $\Delta z=2$\AA \ , for example, $z_1=3$\AA, $z_2=5$\AA, ... $z_8=17$\AA. At each $z_m$, the related local correlation functions, i.e., ISF and DSF are reported as well, see in the next subsection \ref{CDmethod}. Utilizing the above mentioned decay distance $\zeta_{\hat{n}}$, together with the $\tau_{\hat{n}}(z_m)$ profile, we define the density relaxation time of the CMI melt, $\tau_{\hat{n},\mathrm{I}}=\tau_{\hat{n}}(z_m)$, $j$ takes the integer value of $[\lceil \frac{\zeta_{\hat{n}}-3.0}{\Delta z} \rceil +1]$ ($\lceil ... \rceil$ denotes the ceiling function). The values of $\tau_{\hat{n},\mathrm{I}}$ for the three Fe BCC CMIs are employed in the new validation of the TDGL theory in predicting the kinetic coefficients and anisotropy (see Appendix \ref{sec:ap2}). Considering the cost of computing resources, we take 1000 subaverage samples in determining the magnitude and statistical error for the data points which are defined as $\tau_{\hat{n},\mathrm{I}}$, while the rest of the data point along the $\tau_{\hat{n}}(z_m)$ profile use 200 subaverage sample. 

In addition to the above two types, the coarse-grained diffusion coefficient profiles, $D_{\hat{n}}(z)$ for the three Fe CMIs are also calculated, following the detailed method described in Refs.\onlinecite{Davidchack98,Yang12}. The anisotropy among three orientations will be discussed through the comparison between the interfacial $D_{\hat{n}}(z)$ and the $\tau_{\hat{n}}(z_m)$ data.

\subsection{Calculation of collective dynamical properties}
\label{CDmethod}

\subsubsection{Bulk melt phase}
For the bulk melt system, the collective dynamics of the liquid can be resolved by determining the cross-correlations of the particle coordinates, i.e., the ISF $F_\mathrm{M}(|\vec{k}|, t)$ and the DSF $S_\mathrm{M}(|\vec{k}|, \omega)$ defined as 
\begin{equation}
F_\mathrm{M}(|\vec{k}|,t)=\frac{1}{N}\langle \sum^N_{i=1}\sum^N_{j=1}\mathrm{exp}\lbrack -i\vec{k}\cdot(\vec{r}_i(t)-\vec{r}_j(0))\rbrack \rangle,
\label{FM}
\end{equation}

\begin{equation}
S_\mathrm{M}(|\vec{k}|,\omega )=\frac{1}{2\pi }\int^{+\infty }_{-\infty }F_\mathrm{M}(|\vec{k}|,t)\mathrm{exp}(i\omega t )\mathrm{d}t.
\label{SM}
\end{equation}

The calculations in Eq.~\ref{FM} and Eq.~\ref{SM} count all $N$ atoms in the entire simulation cell. The characteristic time-scale (or the density relaxation time) for the bulk melt phase, under a specific wavenumber, $\tau_\mathrm{M}(|\vec{k}|)$, can be calculated from the inverse half-width of DSF.

\subsubsection{Local method by del Rio et al.\cite{Rio20}}

\begin{figure}[!htb]
\centering
\includegraphics[width=0.46\textwidth]{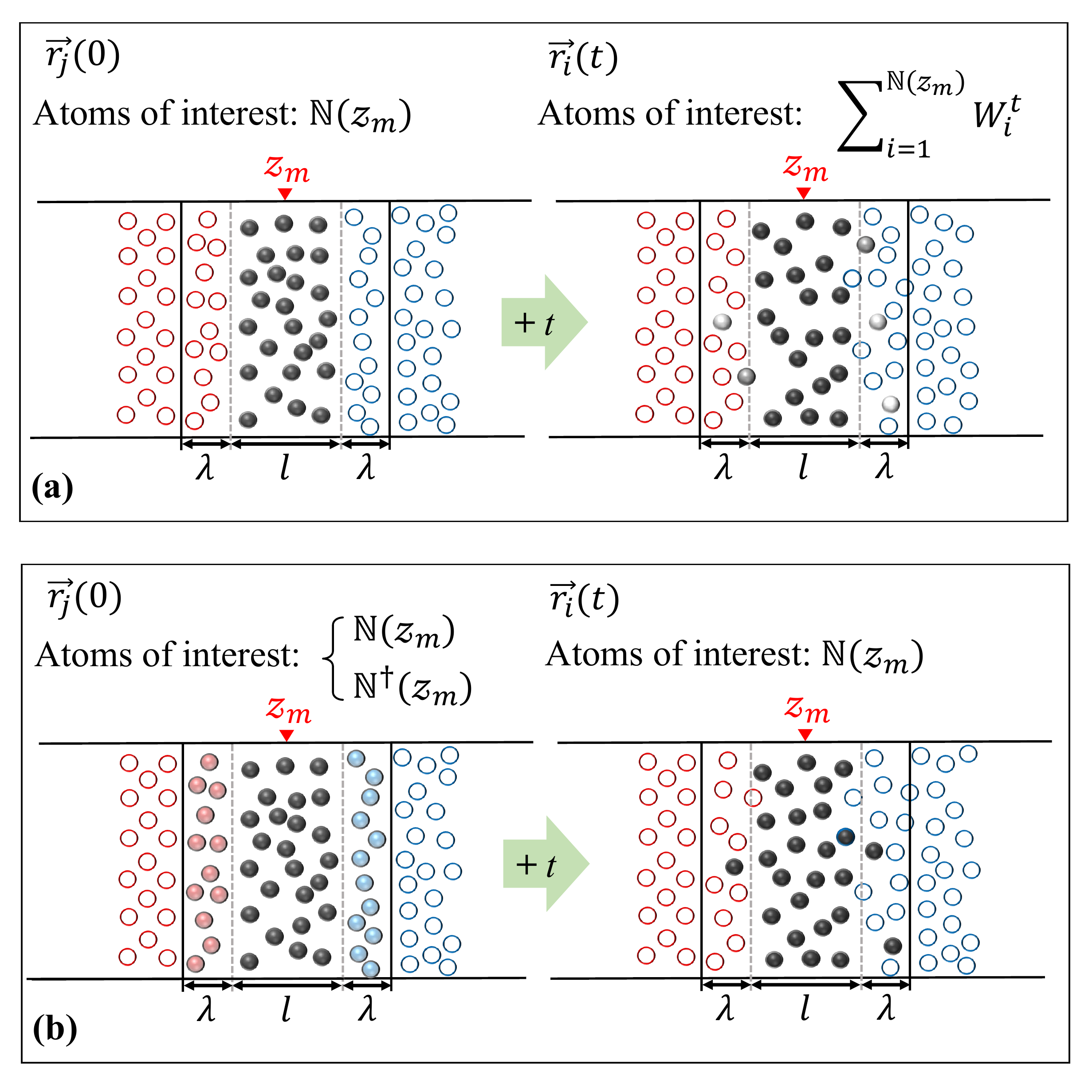}
\caption{Schematic diagrams illustrating the calculation of local ISFs for a slab region ($l$ thickness) at $z=z_m$. The local particle numbers of interest at time $t$ and 0 are listed. $\lambda$ is the thickness of the two adjacent transition regions next to the region of interest. The red and blue open circles represent the solid-like and liquid-like particles surrounding the  atoms of interest (filled symbols). (a) Method proposed in Ref.\onlinecite{Rio20}, i.e., Eq.\ref{Rio-ISF} and Eq.\ref{Rio-W}. The fading color in the filled symbols on the right side depicts the magnitude of their affiliated weight functions $W^t_i$. (b) The method proposed in the current study, Eq.\ref{Zhang-ISF}. More atoms ($\mathbb{N}^{\dagger}(z_m)$) than than are present in the region of interest ($\mathbb{N}(z_m)$) are included at $t=0$ to take into account the possible impact of the adjacent ordering environment on the local collective dynamic properties of interest.}
\label{fig2}
\end{figure}

In 2020, del Rio et al. introduced an atomistic simulation-based methodology for calculating the local collective dynamical properties for a liquid at different depths of the metal within the liquid-vapor interface\cite{Rio20}. As illustrated in Fig.~\ref{fig2}(a), for a slab region of interest with a thickness $l$, with its center position located at $z=z_m$, they proposed that the ISF for this local region is calculated as,

\begin{footnotesize}
\begin{equation}
F(|\vec{k}|,t,z_m)=
\left\langle \frac{
\sum^{\mathbb{N}(z_m)}_{i=1}\sum^{\mathbb{N}(z_m)}_{j=1}W_i^{t}W_j^{0}\exp[-i\vec{k}\cdot(\vec{r}_i(t)-\vec{r}_j(0))]}{\sqrt{\sum^{\mathbb{N}(z_m)}_{i=1}\sum^{\mathbb{N}(z_m)}_{j=1}W_i^{t}W_j^{0}}} \right\rangle,
\label{Rio-ISF}
\end{equation}
\end{footnotesize}
\noindent
in which only the $\mathbb{N}(z_m)$ particles inside the region of interest were included in sampling trajectories. Because liquid particles can diffuse in and out of the region during the delay time window $t$, an atomic weight function $W^t_i$ for particle $i$ was introduced to take this into account,

\begin{footnotesize}
\begin{equation}
W_i^t=\begin{cases} 
0 & \textbf{if  }  z_i(t)<z_m-\frac{l}{2}-\lambda\\
\frac{1}{2}+\frac{1}{2}\cos [\frac{z_i(t)-(z_m-\frac{l}{2})}{\lambda} \pi] & \textbf {if } z_m-\frac{l}{2}-\lambda<z_i(t)<z_m-\frac{l}{2} \\
1 & \textbf{if } z_m-\frac{l}{2}<z_i(t)<z_m+\frac{l}{2} \\
\frac{1}{2}+\frac{1}{2}\cos[\frac{z_i(t)-(z_m+\frac{l}{2})}{\lambda} \pi] & \textbf{if } z_m+\frac{l}{2}<z_i(t)<z_m+\frac{l}{2}+\lambda \\
0 & \textbf{if } z_i(t)>z_m+\frac{l}{2}+\lambda
\label{Rio-W}
\end{cases}
\end{equation}
\end{footnotesize}
where $\lambda$ is the thickness of the transition region in which $W_i^t$ function  decays smoothly to 0. With this weight function, the number of particles changes from $\mathbb{N}(z_m)$ (at $t=0$, $W_j^{0}$=1) to $\sum^{\mathbb{N}(z_m)}_{i=1}W_i^{t}$, after a delay time of $t$. Accordingly, the DSF for this  slab region at $z=z_m$ is expressed as follows,
\begin{equation}
S(|\vec{k}|,\omega,z_m)=\frac{1}{2\pi }\int^{+\infty }_{-\infty }F(|\vec{k}|,t,z_m)\mathrm{exp}(i\omega t )\mathrm{d}t.
\label{Rio-S}
\end{equation}

This local method proposed by del Rio et al. has been proved to be feasible in well-reproducing inelastic X-ray scattering experimental results\cite{Reichert07} of the local liquids (a 4 nm depth probe) at the liquid-vapor interface. In the current study, we will apply this method to the crystal-melt interfaces and compare it with our method (see below) to characterize the collective dynamical properties and predict  CMI kinetic coefficients.

\subsubsection{Local method for CMI system}

For the crystal-melt interfaces, the interfacial liquids can show an evident lateral (in-plane) ordering in addition to  the usual layer ordering normal to the interfacial plane \cite{Kaplan06}. Therefore, it is necessary to include the possible impact of the adjacent ordering environment on  local collective dynamical properties of interest. As illustrated in the schematic diagram in Fig.~\ref{fig2}(b), we propose an algorithm for calculating the local collective dynamics by including the possible impact of the inhomogeneous structural ordering environment. At the time origin $t=0$, coordinates of both $\mathbb{N}(z_m)$ atoms in the region of interest and the atoms in its two neighboring transition regions (thickness $\lambda$) are recorded, yielding a total of $\mathbb{N}^{\dagger}(z_m)$ atoms. This information is employed in the sampling average of the time-spatial inter-correlation between the $\mathbb{N}(z_m)$ atoms at time $t$ and the $\mathbb{N}^{\dagger}(z_m)$ atoms at the time origin, see Eq.\ref{Zhang-ISF}. Accordingly, the DSF for the  local region  at $z=z_m$ is expressed in Eq.\ref{Zhang-S}. The superscript ``$\dagger$'' in $F^{\dagger}(|\vec{k}|,t,z_m)$, $S^{\dagger}(|\vec{k}|,\omega,z_m)$, and other quantities in the following results is to distinguish from those predicted using the method of del Rio et al.

\begin{footnotesize}
\begin{equation}
F^{\dagger}(|\vec{k}|,t,z_m)=\frac{1}{\mathbb{N}(z_m)}\langle \sum^{\mathbb{N}(z_m)}_{i=1}\sum^{\mathbb{N}^{\dagger}(z_m)}_{j=1}\mathrm{exp}\lbrack -i\vec{k}\cdot(\vec{r}_i(t)-\vec{r}_j(0))\rbrack \rangle,
\label{Zhang-ISF}
\end{equation}

\begin{equation}
S^{\dagger}(|\vec{k}|,\omega,z_m)=\frac{1}{2\pi }\int^{+\infty }_{-\infty }F^{\dagger}(|\vec{k}|,t,z_m)\mathrm{exp}(i\omega t )\mathrm{d}t.
\label{Zhang-S}
\end{equation}
\end{footnotesize}

\begin{figure}[!htb]
\centering
\includegraphics[width=0.45\textwidth]{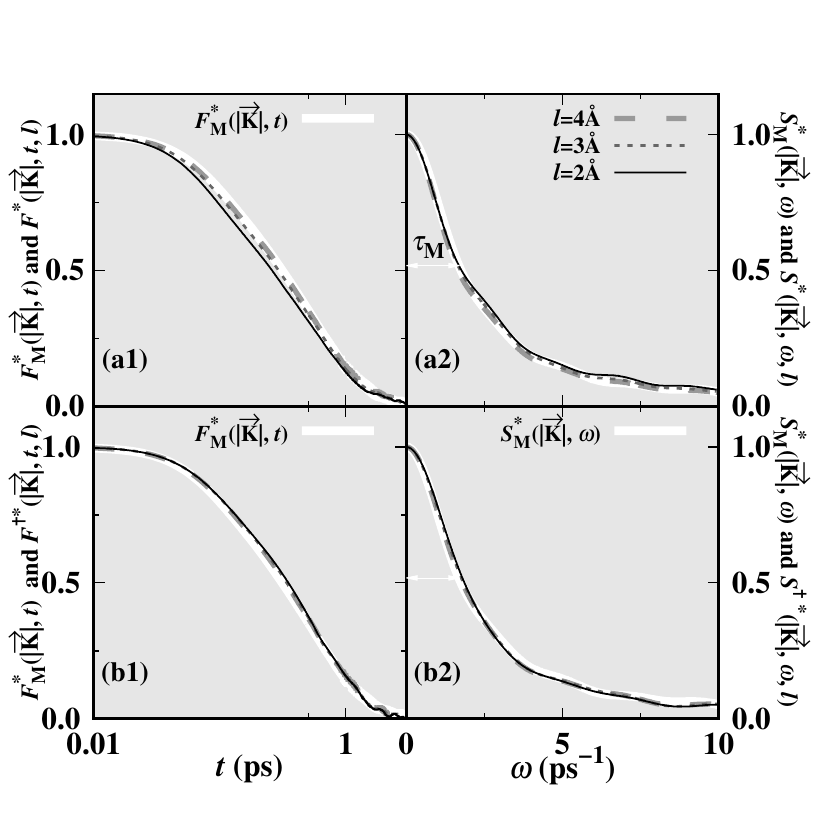}
\caption{Validity check for the two local methods (the local method by del Rio et al. panel (a1-a2); our local method panel (b1-b2)) for the determination of the collective dynamical properties (ISFs and DSFs) for the bulk Fe melt at $T=T_\mathrm{m}$, represented with thinner grey lines. Three different  regions with thickness $l=2$\AA, \ 3\AA, \ and 4\AA, are examined together with the results using the bulk method (Eq.~\ref{FM} and Eq.~\ref{SM}), represented with the white thick solid line. The data are normalized by the time origin values (ISF) or the zero frequency values (DSF), labeled with a superscript ``$*$''.}
\label{fig3}
\end{figure}
 
The results shown in  Fig.~\ref{fig3} suggest that the collective dynamical properties of the bulk liquid can be correctly calculated within both the two local methods using only atom information within a finite thickness region of interest. It is found that when the region thickness $l$ is greater than 4\AA \, the collective dynamical properties (ISFs and DSFs) determined with the local methods are more likely to be consistent with the bulk data. In what follows, the collective dynamical properties are calculated using the local methods with a choice of $l=2\lambda=5.85\AA$, around twice the lattice parameter of BCC Fe at $T=T_\mathrm{m}$. In this study, because our goal is to predict the solidification kinetics, only one specific wavenumber is studied, namely, the value of $|\vec {\mathrm{K}}|$ corresponding to the  shortest non-zero reciprocal lattice vectors (12 degenerate vectors in all) of the BCC Fe at $T=T_\mathrm{m}$ - see Appendix~\ref{sec:ap1} for details.

\subsection{Calculation of the CMI kinetic coefficients}

The values of the relaxation times $\tau_{\hat{n},\mathrm{I}}$ and $\tau^{\dagger}_{\hat{n},\mathrm{I}}$ obtained using the two local methods for the collective dynamics are employed in implementing the TDGL theory (see Appendix~\ref{sec:ap2}) to revisit the kinetic anisotropies and the previously reported discrepancy between the simulation and theory predictions of the CMI $\mu$.

\section{Results and Discussion}

\begin{figure}[!htb]
\centering
\includegraphics[width=0.48\textwidth]{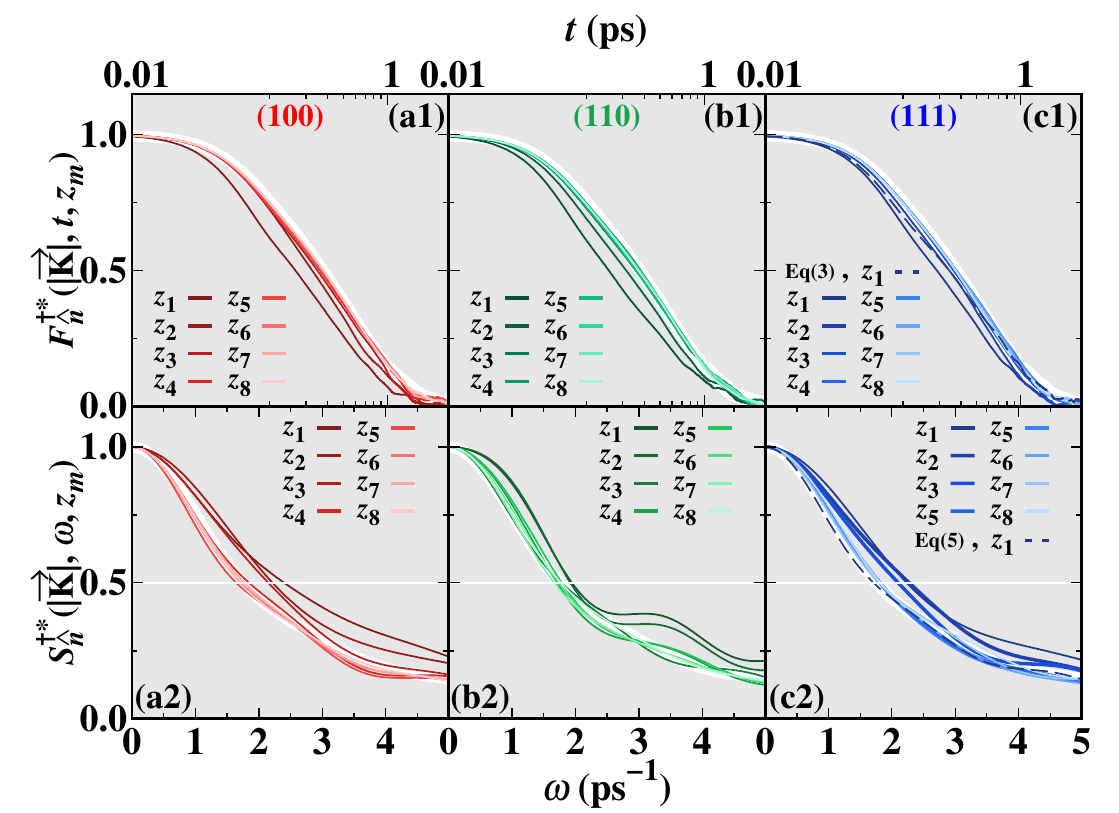}
\caption{The normalized local ISFs (a1-c1) and DSFs (a2-c2) at the principal reciprocal lattice vector, $|\vec {\mathrm{K}}|$, for the different regions of interest located at eight $z_m$ positions of the BCC Fe (100) (red), (110) (green) and (111) (blue) CMIs at $T_\mathrm{m}=1772$ K. The dashed line in (c1) and (c2) shows a local collective dynamics data using the local method by del Rio et al. at $z_1$, while the data represented with solid lines are determined by our local method (Eq.\ref{Zhang-ISF} and Eq.\ref{Zhang-S}). The bulk melt data (white line) is as shown in Fig.\ref{fig3}.}
\label{fig4}
\end{figure}
 
\begin{figure}[!htb]
\centering
 \includegraphics[width=0.45\textwidth]{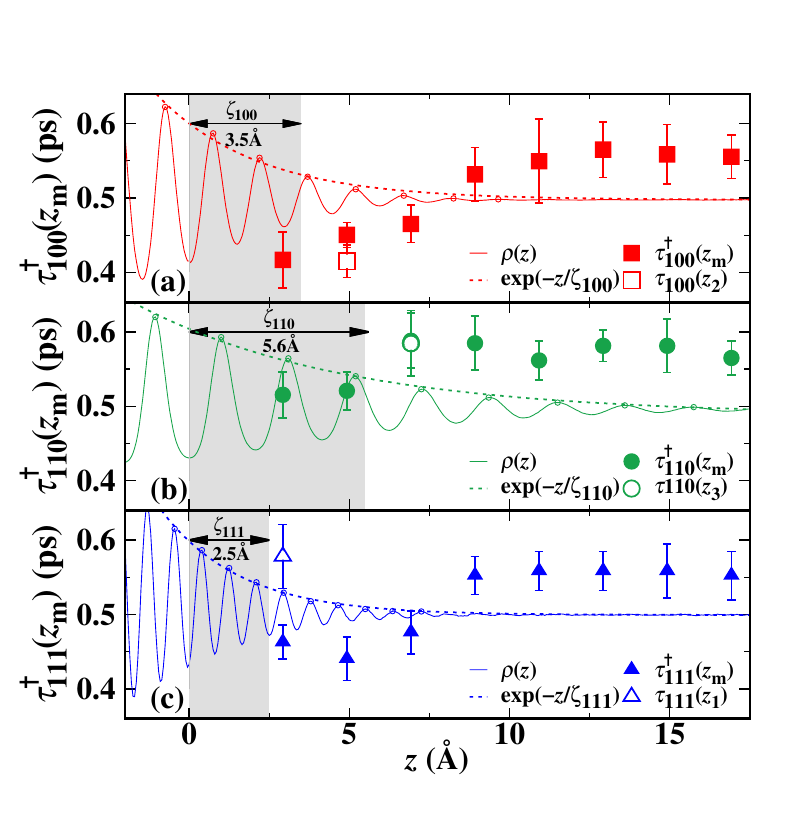}
 \caption{Local density relaxation time profiles of the three BCC Fe CMIs, $\tau^{\dagger}_{\hat{n}}(z_m)$, $\hat{n}=$ (100) (a), (110) (b), and (111) (c). $z=0$ corresponds to the position of the CMI, $z>0$ for melt, $z<0$ for crystal. The fine-grained particle number density profiles $\rho(z)$ together with the exponential fits to the stratification decays are plotted as a guide to the eye. The decay lengths $\zeta_{\hat{n}}$ (shaded area) define the density relaxation times of the CMI melts, $\tau^{\dagger}_{{\hat{n}},\mathrm{I}}$ (our method) and $\tau_{{\hat{n}},\mathrm{I}}$ (del Rio et al. method) represented with an open symbol in each panel.}
\label{fig5}
\end{figure}

Fig.~\ref{fig4} shows the ISF and DSF data for the local liquids at different regions of interest within the three equilibrium BCC Fe CMIs. For all three CMIs, the ISF and DSF results for the interfacial liquids show a faster dynamic response decay than the bulk data.
In other words, the density relaxations in the CMI liquids occur on time scales shorter than those in the bulk melt phase. The density relaxation times (inverse half-width of the DSF) are extracted from the DSFs and are plotted in Fig.\ref{fig5}.

The ISF and DSF data do not show strongly damped oscillations in addition to their nearly exponential decay, as we are treating a relatively large wavenumber, $\vert \vec {\mathrm{K}} \vert=2.985\mathrm{\AA}^{-1}$. Note that, we have confirmed that the local CMI liquid ISF at smaller wavenumbers is well described by a hydrodynamic model expression, in which the ISF is expressed as a sum of a thermal relaxation-related exponential decay function and a sound wave attenuation-related damped oscillation term\cite{Bafile06}. Also, following the idea of del Rio et al.\cite{Rio20}, it would be interesting to carry out an independent study on the wave-vector dependences on the thermodynamics parameters used in such hydrodynamic model, such as the longitudinal viscosity and adiabatic sound velocity, etc. for the local CMI liquids; however, such studies are beyond the scope of the current study.

\begin{table*}[!htb]
\caption{Summary of input parameters used in TDGL theory. Values of $L$, $S$, $T_\mathrm{m}$, $\vert \vec {\mathrm{K}} \vert$ are taken from Ref.\onlinecite{Wu15}. 
The anisotropic factor $A_{\hat{n},\mathrm{I}}$ are calculated using the orientation-dependent interfacial density relaxation time $\tau_{\hat{n},\mathrm{I}}$, Eq.\ref{AnI} in Appendix\ref{sec:ap2}. Error bars represent 95\% confidence intervals on the last digit(s) shown.}
\begin{ruledtabular}
\begin{tabular}{llllllllllllllll}
$L$&$S$&$T_\mathrm{m}$&
$\vert \vec {\mathrm{K}} \vert$&
$A_{100,\mathrm{I}}$&
$A_{110,\mathrm{I}}$&
$A_{111,\mathrm{I}}$&
$A^{\dagger}_{100,\mathrm{I}}$&
$A^{\dagger}_{110,\mathrm{I}}$&
$A^{\dagger}_{111,\mathrm{I}}$&
$\tau_{100,\mathrm{I}}$&
$\tau_{110,\mathrm{I}}$&
$\tau_{111,\mathrm{I}}$&
$\tau^{\dagger}_{100,\mathrm{I}}$ &
$\tau^{\dagger}_{110,\mathrm{I}}$ &
$\tau^{\dagger}_{111,\mathrm{I}}$ \\
eV/atom& -- &K& \AA$^{-1}$ & 
\multicolumn{6}{c} {(                \AA/ps                )} &
\multicolumn{6}{c} {(     ps     )} \\
\hline
0.162&3.02&1772&2.985&
0.21(1)&0.30(2)&0.36(1)&
0.23(1)&0.31(2)&0.29(1)&
0.41(2)&0.58(4)&0.58(4)&
0.45(2)&0.59(4)&0.46(2)
\\
\end{tabular}
\end{ruledtabular}
\label{tab1} 
\end{table*}

\begin{figure}[!htb]
\centering
 \includegraphics[width=0.45\textwidth]{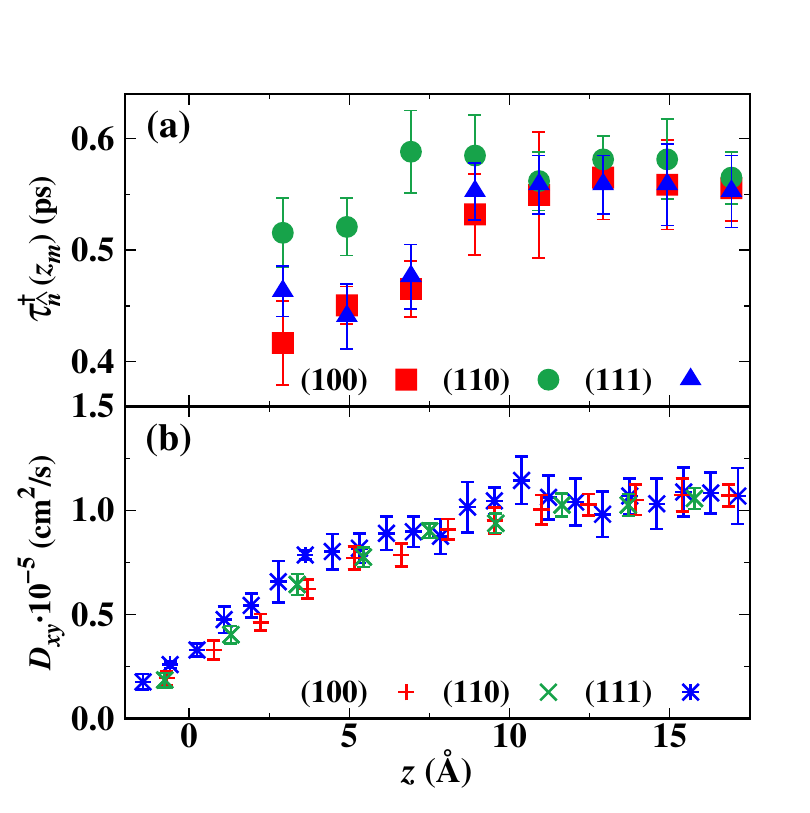}
 \caption{(a) Pronounced anisotropy in the $\tau^{\dagger}_{\hat{n}}(z_m)$ profiles for the three BCC Fe CMIs. (b) Weak anisotropy among the diffusion coefficient profiles of the three CMIs.}
 \label{fig6}
 \end{figure}

Figs.~\ref{fig5}(a) (b) and (c) show the coarse-scale profiles of the density relaxation times $\tau^{\dagger}_{\hat{n}}(z_m)$ across the (100), (110) and (111) Fe CMI, respectively. As the interface is traversed from the melt phase to the crystal phase, the magnitude of  $\tau^{\dagger}_{\hat{n}}(z_m)$ decreases as the CMI positions ($z=0$) is approached. The reduction of the relaxation time in the interfacial region can reach around 10\% to 30\% of the bulk melt phase value ($\tau_{\mathrm{M}}=0.57(5)$ps\cite{Wu15}). The speed up of the collective dynamics in the CMI region could be attributed to the fact that the activation energy of crystallization\cite{Rodrigo20} for the liquid-like particles is lower within the CMI region under the influence of the nearby crystal. Previous studies have suggested, in the bulk melt phase, the liquid density relaxation times are inversely correlated with the self-diffusion coefficient\cite{Cohen87,Loef89}. As shown in Fig.\ref{fig6}, the magnitudes of the interfacial diffusion coefficients decrease as the crystal is approached. Therefore the correlation between the collective dynamics and the self-diffusion dynamics does not hold for the liquid within the CMI region, and the information on the interfacial diffusion coefficients is not useful in interpreting the variation  in $\tau^{\dagger}_{\hat{n}}(z_m)$.

It is interesting to point out the current finding that the speed up in the collective dynamics of the liquid within the CMI region is significantly different from the observation of the slowdown of the collective dynamics observed by Reichert et al\cite{Reichert07}. for the liquid near the molten indium surface. Reichert et al. argued that the slowdown of the collective dynamics could be related to the stratification of the surface liquid. It is known that  pronounced stratification (or layering) effect also exists in the CMI systems. The most conspicuous difference between a CMI and a liquid-vapor interface is the in-plane ordering\cite{Kaplan06}, which might play a dominant role in facilitating the completion of density relaxation processes. However, to the best of the authors' knowledge, few clues that connect the in-plane ordering and the collective dynamics based on inhomogeneous fluid statistical mechanics can be found in the existing literature.
 
Fig.~\ref{fig6}(a) shows the $\tau^{\dagger}_{\hat{n}}(z_m)$ profiles of the three Fe CMIs. The data exhibit pronounced anisotropy, with the local interfacial density relaxation times of the BCC (110) CMI being greater than those of the BCC (100) and (111) CMIs. For the (100) and (111) interfaces, the $\tau^{\dagger}_{\hat{n}}(z_m)$ are identical within the error bars. Despite that, the layer ordering in (110) CMI extends deep into the melt phase with a larger stratification decay length $\zeta_{110}$ than the (100) and (111) CMIs. It has been reported from a BCC crystal wall-fluid simulation study that the in-plane ordering can hardly be found within the interfacial liquid layers in BCC (110) system. The BCC (100) system exhibits in-plane ordering within the first few liquid layers next to the substrate. It is presumed that the reason for the relatively slower local collective dynamics for the BCC (110) CMI is related to the lack of in-plane ordering, which agrees with our inference discussed in the preceding paragraph. 

In contrast to the significant anisotropy seen in Fig.\ref{fig6}(a), the three interfacial diffusion profiles of the three Fe CMIs are nearly identical, see in panel (b) of the Fig.\ref{fig6}. Once again, this comparison suggests the self-diffusion dynamics and the collective dynamics are not strongly correlated in the CMI region, and the diffusion coefficient data does not help understand the solidification kinetic anisotropy.

The method by del Rio et al. (Eq.\ref{Rio-ISF}, Eq.\ref{Rio-S}) has also been employed in determining the ISFs, DSFs, and the $\tau_{\hat{n}}(z_m)$ at the three CMIs. Within the statistical uncertainties, the results produced from the del Rio et al. method for (100) and (110) CMIs are nearly identical to the results produced from our local method. See the open symbols in Fig.\ref{fig5}(a) and (b), for instance. However, for the (111) CMI, the method by del Rio et al. predicts significantly different results, i.e., the collective dynamics of the (111) interfacial liquids  has a magnitude that is similar to that of the bulk melt phase $-$ see the dashed curves Fig.~\ref{fig4}(c1-c2) and the open symbols in Fig.~\ref{fig5}(c). The significant difference between the calculated results for the (111) CMI indicates that there are likely two different microscopic mechanisms that determine the local collective dynamics among the three BCC CMIs. For the BCC (100) and (110) CMIs, where the interlayer spacings for the interfacial liquid layers are larger than 2\AA, the impact of the adjacent ordering environment on the  local collective dynamics is not conclusive. Both the two local methods can fully capture the contributions of the local collective dynamics;  whereas, for the BCC (111) CMI, where the interlayer spacing is only $\sim$30\% of the lattice parameter, the in-plane ordering of a liquid interfacial layer highly relies on inheriting its neighboring lattice motif. The local collective dynamics of the  region of interest could be underestimated due to  reduced consideration of a portion of the local structural ordering.

\begin{table}[!htb]
\caption{Summary of the magnitude and anisotropy of the BCC Fe CMI kinetic coefficients predicted by the non-equilibrium MD simulations\cite{Gao10}, by the TDGL theory employing the bulk melt phase density relaxation time\cite{Wu15}, and by the TDGL theory employing the local interface liquid density relaxation times, which are determined by the local method by del Rio et al., and our local method (labeled with $\dagger$).}
\begin{ruledtabular}
\begin{tabular}{lllll}
&$\mu^{\mathrm{MD}}_{\hat{n}}$&$\mu^{\mathrm{GL}}_{\hat{n},\mathrm{M}}$&$\mu^{\mathrm{GL}}_{\hat{n},\mathrm{I}}$&$\mu^{{\dagger}\mathrm{GL}}_{\hat{n},\mathrm{I}}$\\
\hline
$\mu_{100}$(cm/s/K)& 78(5)        &65(6)  &90(4)&82(4)\\
$\mu_{110}$(cm/s/K)& 62(4)        &60(5)  &59(4)&58(4)\\
$\mu_{111}$(cm/s/K)& 62(2)        &51(5)  &50(2)&63(3)\\
$\mu_{100}/\mu_{110}$ &1.27(16)& 1.08(13) &1.53(12)&1.41(12)\\
$\mu_{100}/\mu_{111}$ &1.26(12)& 1.27(17) &1.80(10)&1.30(9)\\
$\mu_{110}/\mu_{111}$ &0.99(8)  & 1.18(15) &1.18(9)&0.92(7)\\
\end{tabular}
\end{ruledtabular}
\label{tab2}
\end{table}

The data listed in the 2nd column of Table~\ref{tab2} are the previous TDGL predictions ($\mu^{\mathrm{GL}}_{\hat{n},\mathrm{M}}$) by Wu et al\cite{Wu15}. using the bulk liquid density relaxation time $\tau_\mathrm{M}$, which underestimates the $\mu^{\mathrm{MD}}$ by about 15\% $\sim$ 20\% for the (100) and (111) CMIs, resulting in the ratios $\mu_{100}/\mu_{110}$ and $\mu_{110}/\mu_{111}$ departs from the MD values.

With the interfacial density relaxation times (listed in Table~\ref{tab1}), we revisit the TDGL theory prediction of the $\mu_{\hat{n}}$ and the kinetic anisotropy for the Fe BCC CMIs, see in Table~\ref{tab2}. The data listed in the fourth column in Table~\ref{tab2} use the interfacial density relaxation times calculated by our method, yielding overall excellent agreement with both the magnitudes and the kinetic anisotropies of the kinetic coefficients. The 95\% confidence error bars of $\mu_{\hat{n}}^{\mathrm{MD}}$ and $\mu^{{\dagger}\mathrm{GL}}_{\hat{n},\mathrm{I}}$, as well as their anisotropies all overlap. Because the $\tau_{111,\mathrm{I}}$ of the BCC (111) interfacial liquids determined via the method by del Rio et al. is significantly larger than $\tau^{\dagger}_{111,\mathrm{I}}$, TDGL predictions listed in the 3rd column of Table~\ref{tab2} represent an even larger discrepancy scenario in comparing with the 2nd column.

The comparison of the second and fourth columns suggests that a primary contributing factor to this significant improvement is that the current study uses the anisotropic collective dynamical properties of the local CMI liquids, calculated with the local method including the impact of the adjacent ordering environment. With this improvement in theoretical prediction, new knowledge relating to the origin of the solidification kinetic anisotropy can be obtained. It is due to the orientation-dependent interfacial collective dynamics property ($\tau^{\dagger}_{\hat{n},\mathrm{I}}$) further carving the kinetic anisotropy based on the kinetic anisotropy with only focus on the generic entropic-driven particle packing, i.e., the 2nd column of Table~\ref{tab2}, which uses an orientation-independent bulk melt value of $\tau_{\mathrm{M}}$. 

\section{conclusion}

In the current study, the local collective dynamics of the interfacial liquid regions at equilibrium BCC Fe (100), (110), and (111) CMIs are calculated based on atomistic simulations. Two methods of determining the local collective dynamics of the liquids in an interfacial region of interest with a finite thickness are employed, specifically, the method by del Rio and Gonz\'{a}lez proposed in 2020\cite{Rio20} and a new method considering the possible impact of the adjacent ordered environment. The spatial variations of the ISFs, DSFs, and the density relaxation times (at a specific wavenumber $\vert \vec {\mathrm{K}} \vert$) spanning over around 20 \AA \ of the three orientations are presented. It is found that local collective dynamics speed up in the CMIs region, which is significantly different from the observed slowing down of the local collective dynamics at the liquid-vapor interface\cite{Rio20}.

In addition, the speed up in the local collective dynamics exhibit pronounced orientational anisotropy among the three BCC CMIs. We presume that the in-plane ordering may play a dominant role in facilitating the collective dynamics and their anisotropy in CMI regions, whereas the local mass transport (self-diffusion coefficients) seems irrelevant. By including the density relaxation times for the interfacial liquids (calculated by our method), we demonstrate that the TDGL theory can accurately predict both the magnitude and the kinetic anisotropy of the CMI kinetic coefficients measured from the non-equilibrium MD simulations, $\mu^{\mathrm{MD}}_{\hat{n}}$, yielding a reduction of the discrepancy reported in its previous validation\cite{Wu15}. In contrast, by including the density relaxation times calculated by the method of del Rio et al., unsatisfactory TDGL prediction of the kinetic coefficient for the CMI with relatively narrow interlayer spacing is found, due to the insufficient counting of the nearby structural ordering impact in their method.

Overall, there are currently no inhomogeneous fluid statistical mechanical theories for quantitatively predicting the specific variations (e.g., slowing down at the liquid-vapor interface, speeding up at CMIs, and so on) and their crystalline anisotropies. Half of the hypotheses proposed in our previous work\cite{Wang22}, i.e., the density relaxation times for the interface melt phases should be anisotropic and material-dependent, have been verified in the current study. More efforts are suggested to resolve the collective dynamical anisotropies of the CMI liquids over various BCC elemental materials. We hope that additional calculated data would benefit the advancement of the predictive theory for the collective dynamics of the local interfacial liquids.

This study gives us an important revelation: the interfacial liquid density relaxation time is a crucial parameter with the potential for modulation. In-depth investigations on this local dynamical property (such as its material-, composition- and structure- dependencies) could benefit the quantitative understanding of recently reported anomalously low solidification rates in glass-forming and high entropy alloy systems\cite{Tang13,Zhang20} and provide insights for steering the solidification kinetics utilizing external electric or magnetic fields\cite{Xu20,Ren14}.

\begin{acknowledgments}
YY acknowledges the Chinese National Science Foundation (Grant No. 11874147), the Natural Science Foundation of Chongqing, China (Grant No. cstc2021jcyj-msxmX1144), Open Project of State Key Laboratory of Advanced Special Steel, Shanghai Key Laboratory of Advanced Ferrometallurgy, Shanghai University (SKLASS 2021-10), the Science and Technology Commission of Shanghai Municipality (No. 19DZ2270200, 20511107700) and the State Key Laboratory of Solidification Processing in NWPU (Grant No. SKLSP202105).
\end{acknowledgments}

\appendix

\section{TDGL theory of  CMI  kinetic coefficients}
\label{sec:ap1}

For pure substances in the near-equilibrium linear-response temperature region, the CMI kinetic coefficients $\mu_{\hat{n}}$ with given interface orientation $\hat{n}$, are defined as, 
\begin{equation}
\mu_{\hat{n}}=\frac{V_{\hat{n}}}{\Delta T},
\end{equation}
where, $V_{\hat{n}}$ is the steady state interface velocity, $\Delta T=T_\mathrm{m}-T$ is the interface undercooling and $T_\mathrm{m}$ is crystal-melt coexistence temperature (melting point). 

The TDGL theories for the near-equilibrium BCC and FCC CMI kinetics, developed by Wu et al\cite{Wu15}. and Xu et al.\cite{Xu20}, are  state-of-the-art theories for the quantitative prediction of  both the magnitude and  anisotropy of the CMI kinetic coefficients, $\mu_{\hat{n}}$. In these theories, the excess free energy of the non-equilibrium CMI is represented by the sum of the thermodynamic driving force of the solidification and the free-energy functional form of the equilibrium CMI derived initially from classical density functional theory (cDFT).  The Ginzburg-Landau expansion of the free-energy density as a function of the relevant order parameters (defined as the amplitudes of density waves corresponding to the crystal reciprocal lattice vectors, RVLs for short) is employed in the formalism of the CMI excess free energy. Thusly constructed, the TDGL equation describes the energy dissipation and the dynamical evolution of the order parameter field during solidification. Here, the time derivative of the order parameter (or the ordering flux corresponding to different RLV sets) is assumed to be proportional to the local gradient of the free energy functional concerning the order parameter (or their driving forces) via a microscopic time scale related to density wave relaxation.

For the elemental BCC CMI system, the TDGL theory produces an analytical expression for the CMI kinetic coefficients,
\begin{equation}
\mu^\mathrm{GL}_{\hat{n},\mathrm{M}}=\frac{LS}{k_\mathrm{B} T^2_\mathrm{m} A_{\hat{n},\mathrm{M}}},
\label{MunM}
\end{equation}
 in which the superscript ``GL'' stands for the TDGL theory prediction,  $k_\mathrm{B}$ is Boltzmann's constant, $L$ is the latent heat at $T_\mathrm{m}$ and $S$ is the static structure factor of the melt phase at wave number of $|\vec {\mathrm{K}}|$, where $\vec {\mathrm{K}}$ represents the shortest 12 non-zero reciprocal lattice vectors of the BCC crystal at $T=T_\mathrm{m}$.The subscript ``M'' here and in what follows denotes  data obtained from the bulk melt phase. 
In addition, the anisotropic factor $A_{\hat{n},\mathrm{M}}$ has the following explicit analytical expression from Ref.\onlinecite{Wu15},
\begin{equation}
A_{\hat{n},\mathrm{M}}=\int  \tau_{\mathrm{M}}  \sum^{12}_{\vec {\mathrm{K}},i=1}\left[\frac{\mathrm{d}u_i(z)}{\mathrm{d}z}\right]^2 \mathrm{d}z ,
\label{AnM}
\end{equation}
in which $\tau_{\mathrm{M}}$ is the microscopic time scale (density wave relaxation time) corresponding to the inverse half-width of the dynamical structure factor $S_{\mathrm{M}}(\vert \vec {\mathrm{K}} \vert,\omega)$ and $u_i(z)$ represents the spatial distribution of the GL order parameters describing density wave amplitudes corresponding to the $i$th RLV $\vec {\mathrm{K}}$, along the CMI normal direction $z$. 

Wu et al. have calculated the spatial integration of the derivative of $u_i(z)$ for the three BCC Fe CMIs studied here. 
The details of these calculations are shown in $u_i(z)$ Table~\ref{tab-SISG}. The data therein are divided into several categories, 
i.e., $u_a$, $u_b$ and/or $u_c$, which are introduced here. For example, there are four RLVs that have the same symmetry with respect t
o $\hat{n}=(100)$ and, thus, have identical values of $(\hat{\mathrm{K}}\cdot\hat{n})^2$. These four order parameters are assigned to $u_a$ 
category. The key ingredients in Eq.~\ref{AnM}, namely, the spatial integrations of the square-gradient terms of the $u_i(z)$ profiles (SISG), 
are listed in the Table~\ref{tab-SISG}, as well. Note that the data Table~\ref{tab-SISG} are taken from Ref.\onlinecite{Wu15} by Wu et al.

\begin{table}
\caption{Categories of the GL order parameters, as well as the corresponding spatial integrations of the square-gradient terms of the $u_i(z)$ profiles (SISG) in Eq.~\ref{AnM}. Note that the $z$ coordinates and the GL order parameters in the integrands are rescaled by the liquid correlation length $\xi_\mathrm{M}$ and the bulk crystal GL order parameter value $u_\mathrm{C}$, respectively. The data listed in this table are taken from Ref.~\onlinecite{Wu15}.}
\begin{ruledtabular}
\begin{tabular}{lccccccc}
&\multicolumn{2}{c}{$\hat{n}=(100)$}&\multicolumn{3}{c}{$\hat{n}=(110)$}&\multicolumn{2}{c}{$\hat{n}=(111)$}\\
\hline
OP catogory&$u_a$&$u_b$&$u_a$&$u_b$&$u_c$&$u_a$&$u_b$\\
Numbers of $\vec{\mathrm{K}}$&4&8&2&8&2&6&6\\
$(\hat{\mathrm{K}}\cdot\hat{n})^2$&0&1/2&0&1/4&1&0&2/3\\
$\int \mathrm{d}\tilde{z}(\frac{\mathrm{d}\tilde{u}_i}{\mathrm{d}\tilde{z}})^2$ &0.37&0.28&0.45&0.33&0.23&0.52&0.27\\

\end{tabular}
\end{ruledtabular}
\label{tab-SISG}
\end{table}

A value of 0.57(5) ps for $\tau_{\mathrm{M}}$ of the bulk molten Fe under $T=T_\mathrm{m}$ has been calculated by Wu et al. These data,  as well as, the remaining parameters in Eqs.~\ref{MunM} and~\ref{AnM}, which are listed in Table~\ref{tab1} and Table~\ref{tab-SISG}, reproduce the TDGL predictions of $\mu^\mathrm{GL}_{\hat{n},\mathrm{M}}$ in  Ref.\onlinecite{Wu15}, see in Table~\ref{tab2}.

Because  the constant $\tau_{\mathrm{M}}$ is employed in all three CMI orientations, the kinetic coefficient anisotropy using Eq.~\ref{MunM} and Eq.~\ref{AnM} is governed simply by the combination scenarios among different SISG sets. For example, a sizable anisotropy between $\mu^{\mathrm{GL}}_{111,\mathrm{M}}$ and $\mu^{\mathrm{GL}}_{100,\mathrm{M}}$ (or $\mu^{\mathrm{GL}}_{110,\mathrm{M}}$) is predicted due to the fact that the former has six larger value SISG terms than other two orientations. Similarly, near identical $\mu^{\mathrm{GL}}_{100,\mathrm{M}}$ and $\mu^{\mathrm{GL}}_{110,\mathrm{M}}$ can be expected.

\section{Validation of the TDGL theory using local density relaxation times}
\label{sec:ap2}
In this study, we employ the local density relaxation times at three CMI orientations, determined using both the method by del Rio et al\cite{Rio20}. and the new method outlined in this work, for revisiting the TDGL theory prediction of the kinetic coefficients for three different BCC Fe CMI orientations. Here, the density relaxation times of the local interfacial liquids differ from the corresponding value in the bulk melt phase and are anisotropic, and we use $\tau_{\hat{n},\mathrm{I}}$ to distinguish from the bulk melt value $\tau_{\mathrm{M}}$, the subscript ``$\hat{n}$'' and ``I'' denote that the data obtained from the interface melt and is orientation-dependent.

In this new validation, the TDGL analytical expressions for the CMI kinetic coefficients and the anisotropic factor are written as,
\begin{equation}
\mu^\mathrm{GL}_{\hat{n},\mathrm{I}}=\frac{LS}{k_\mathrm{B} T^2_\mathrm{m} A_{\hat{n},\mathrm{I}}},
\label{MunI}
\end{equation}
\begin{equation}
A_{\hat{n},\mathrm{I}}=\int  \tau_{\hat{n},\mathrm{I}}  \sum^{12}_{\vec {\mathrm{K}},i=1}\left[\frac{\mathrm{d}u_i(z)}{\mathrm{d}z}\right]^2 \mathrm{d}z .
\label{AnI}
\end{equation}

The SISG terms in Eq.~\ref{AnI} are as same as in Eq.~\ref{AnM}. Xu et al \cite{Xu20}. offered evidence that the magnitudes of the SISG terms depend only on the crystalline orientation and the crystal structure (or RLVs) and is ascribed to the generic entropy-driven particle packing. The detailed calculation methods of the $\tau_{\hat{n},\mathrm{I}}$ can be found in the main text. The anisotropic $\tau_{\hat{n},\mathrm{I}}$ used in Eqs.~\ref{MunI} and~\ref{AnI} play a role in further elucidating  the magnitudes of the kinetic coefficients and the kinetic anisotropy, in addition to the generic entropy-driven particle packing.

\bibliography{ref}
\end{document}